\documentstyle[preprint,aps]{revtex}
\begin{document}
\preprint{\begin{tabular}{l}
\hbox to\hsize{ \hfill KIAS-P98027}\\[-3mm]
\hbox to\hsize{ \hfill KAIST-12/98}\\[-3mm]
\hbox to\hsize{1998 October \hfill SNUTP 97-067}\\[-3mm]
\hbox to\hsize{ \hfill Brown-HET-1083}\\[5mm] \end{tabular} }

\title{ Mass Matrix Ansatz for Degenerate Neutrinos Consistent with
Solar and Atmospheric neutrino Data}
\draft
\author{Kyungsik Kang$^{(a),(b)}$, Sin Kyu Kang$^{(b)}$, Jihn E. Kim$^{(b),(c)}$
       and Pyungwon Ko$^{(d)}$ }
\address{$^{(a)}$ Department of Physics, Brown University, Providence,
Rhode Island 02912, USA}
\address{$^{(b)}$ School of Physics, Korea Institute for Advanced Study, Seoul 130-012, 
Korea}
\address{$^{(c)}$ Departmen of Physics, Seoul National University, Seoul , Korea}
\address{$^{(d)}$ Department of Physics, KAIST, Taejon 305-701, Korea}
\date{\today}
\maketitle
\tighten
\begin{abstract}
We  suggest mass matrices for neutrinos and charged leptons that
can explain solar and atmospheric neutrino data.
The resulting flavor mixing matrix $V_{\nu}$ has a property that
$(V_{\nu})_{13}=0$, thus making $\nu_e \leftrightarrow \nu_{\mu}$
and $\nu_{\mu} \leftrightarrow \nu_{\tau}$ oscillations to be
effectively a two-channel problem.
Phenomenological consequences of the lepton mass matrix 
ansatze are  consistent with the current data 
on various type of  neutrino oscillation experiments except the LSND 
measurement. 
Three neutrinos, being almost degenerate with $\sum | m_{\nu_i} | \lesssim 1 $ 
eV, can be a part of hot dark matter without any conflict with the 
constraint from neutrinoless double beta decay experiments.
The $\nu_{\mu} \leftrightarrow \nu_{\tau}$ oscillation,
$\sin^2{2\theta_{\mu \tau}}$, is predicted to be 0.86-0.97 with
$\Delta m^2_{\mu \tau}\simeq 2 \times 10^{-3}~\mbox{eV}^2$, which
is consistent with the atmospheric neutrino data and can be tested
further at the planned MINOS and K2K experiments searching for
$\nu_{\mu} \rightarrow \nu_{\tau}$ oscillation.
\end{abstract}
%
\newpage
\narrowtext


\section{Introduction}
In this paper, we suggest a specific form of neutrino and charged lepton
mass matrices
that follow from permutation symmetry and the quark-charged lepton symmetry,
and show that an almost degenerate scenario among three 
flavor neutrinos \cite{kkkk,degenerate} can explain solar and atmospheric 
neutrino data
following from the standpoint of the mass matrix ansatze. 
The parameters of the neutrino mass matrix are constrained by 
the solar and the atmospheric neutrino deficits and the neutrinoless double 
beta decay experiment. Then the 
$\nu_{\mu} \leftrightarrow \nu_{\tau}$ oscillation,
$\sin^2 2\theta_{\mu \tau}$, is predicted to be 0.86--0.97
with $\Delta m_{\mu\tau}^2 \simeq 2\times 10^{-3}~{\rm eV}^2$, 
which is consistent with the atmospheric neutrino data \cite{superk2}.
Three neutrinos, being almost degenerate with $\sum |m_{\nu_i}|\lesssim 1$ eV,
can be a part of hot dark matter of the universe. 

The flavor mixing, the fermion masses, and their hierarchical patterns
still remain to be one of the most fundamental problems in particle 
physics.  As an attempt toward the understanding of the quark mass hierarchy
and the flavor mixing, the quark mass matrix ansatz was introduced 
by Weinberg \cite{wein}.
The key idea is to make the number
of parameters in the mass matrix to be less than the total number of 
flavor mixing parameters, so that there result relations between
mixing parameters and mass eigenvalues.  Some
call it  {\it calculability}.  In particular, 
the Cabibbo angle is calculable in terms of the quark masses 
in this scheme.  Weinberg's idea of {\it calculability} was extended 
for three and more generations by Fritzsch \cite{fritz} and Kang 
{\it et al.} \cite{kang1}.
Since then, the Fritzsch-type mass matrix had attracted a great
deal of attention until the top quark was discovered.
But the Fritzsch texture predicts the top quark mass to be at most about
100 GeV and thus was ruled out \cite{kangt}.

Nevertheless, the Fritzsch type mass matrix is attractive
due to its simplicity. Though its original form is phenomenologically
ruled out, one may want to generalize the Fritzsch texture
by introducing just one more parameter but by maintaining the
calculability  property.    The obvious next move is to increase a nonvanishing 
entry in the Fritzsch-type mass matrix at the $(2,2)$ element. 
Recently, a systematic phenomenological study of such generalized  
mass matrix has been studied by Kang and Kang \cite{kang2}, which is
parametrized by
\begin{eqnarray}
   M_H = \left( \begin{array}{ccc}
               0 & A & 0 \\
               A & D & B \\
               0 & B & C  \end{array} \right).
\end{eqnarray}
The case of $D=0$ reduces to the original Fritzsch type.
As shown in Ref. \cite{kang2}, 
this form can be achieved by successive breaking of the maximal permutation 
symmetry in the mass matrix. Various mass matrix ansatze proposed by 
others \cite{others1} can be identified as special cases of the above form 
by appropriately relating $D$ to $B$.
It has also been shown in \cite{kang2} that the mass matrix (1) with a finite 
range of non-zero relation of $D$  to $B$ can be consistent with experimental 
results including heavy top quark mass, while ruling out several special
$D/B$ ratios considered in \cite{others1}.

Regarding the phenomenological form of the mass matrix
Ramond {\it et al.} \cite{ramond} narrowed down a few years ago possible forms of 
mass matrices having texture zeros at the supersymmetric unification scale.  
Eq.(1) was, of course,  one of the mass matrix patterns considered in Ref.~
\cite{ramond}.  However, they constructed the different patterns of mass 
matrices for the up- and down-quark sectors, whereas Ref. \cite{kang2} 
assumed the same form of mass matrices for both sectors. 
In this paper, we assume the same form of mass matrix for the  charged
lepton sector because  the charged leptons  exhibit a similar hierarchy in mass.

On the other hand, all neutrino masses are zero and lepton numbers
are exactly conserved in the context of the standard model(SM).
However, the current experimental anomalies of 
solar \cite{homestake,gallex,sage,kamioka} and atmospheric 
\cite{kamioka2,superk,soudan,imb} neutrinos lead
us to speculate nearly degenerate but non-zero neutrino masses and mixing
among the three flavors,
as they can be interpreted as originating from the neutrino oscillations.
The deficit of the solar neutrino flux is sometimes
explained economically by the Mikheyev-Smirnov-Wolfenstein (MSW)
effect \cite{msw}.
The ``atmospheric neutrino anomaly" can be interpreted by
the muon neutrino oscillation into other neutrino, possibly, of tau flavor.
The recent CHOOZ experiment \cite{chooz} which is a long baseline experiment 
indicates that one has to invoke a large mixing between the $\nu_{\mu} 
\leftrightarrow \nu_{\tau}$,
which is supported by the more recent result from the Super-Kamiokande
Collaborations \cite{superk2}.
It has been suggested by several authors \cite{kkkk,degenerate} that almost 
degenerate neutrinos are needed to accommodate the solar and atmospheric 
neutrino observations as well as the cosmological constraint that arises 
when we regard neutrinos as  candidates for the hot dark matter within 
the three-flavor framework.

\section{Mass matrix for charged lepton}
Let us start with the new class of mass matrix  Eq.(1) for charged lepton.
Since the matrix $M_H$ contains four independent parameters, one might
think that  the ``{\it calculability}" \cite{kang3} might have been lost.
However, one can make additional ansatz to relate $B$ and $D$ via 
$B=wD$ with  the same ratio parameter $w$ for both the up- and down-quark 
sectors, to maintain the ``{\it calculability}" \cite{kang3}. 

In this paper, we assume the quark-charged lepton symmetry for the mass matrix
so that the matrix form of charged lepton sector is exactly the same as
the new type of quark mass matrix Eq.(1).
Let us diagonalize the mass matrix of the charged lepton sector.
In general, a hermitian matrix $M_H$ can be diagonalized by a single unitary
transformation $U_{L(R)} M_H U^{\dagger}_{L(R)}, $ while
a mass matrix needs a biunitary transformation
$U_LM_HU^{\dagger}_{R}=diag[m_1,m_2,m_3]$ in general. 
Then, we can write 
$U_LM_HU^{\dagger}_L=K\cdot diag[m_1, m_2, m_3]$ where
$K=U_LU^{\dagger}_R$ is a diagonal matrix having the diagonal elements
$\pm 1$ or a phase factor $e^{i\phi}$ in general. 
Since we deal with the real mass matrix in our
problem, $U_L$ is a real matrix and $K$ is real too.
As discussed in Ref.\cite{kang2}, 
because of the empirical mass hierarchy $m_1 \ll m_2 \ll m_3$,
$K=diag[1,-1,1]$ irrespective
of the sign of $D$ and $K=diag[-1,1,1]$ only for positive $D$. 
In view of the hierarchical pattern of the charged lepton masses, it is also
natural to expect that $A < |D| \ll C $, and the case of $K=diag[1,-1,1]$
for positive $D$ can be excluded if the same ratio parameter $w$ 
as that for the quark mass matrices is required.
Then, the parameters $A, B, C$ and $D$ can be expressed in terms of
the charged lepton masses and $w$.
For the other case $K = diag[1,1,-1]$,
the characteristic equation for the mass matrix does not admit any
solution.

{\it The Case I {\rm with} $K=diag[-1, 1, 1]$}: From 
the characteristic equation for the $M_H$, 
the mass matrix $M_H$ can be written by
\begin{eqnarray}
   M_H = \left( \begin{array}{ccc}
               0 & \sqrt{\frac{m_1 m_2 m_3}{m_3-\epsilon}} & 0 \\
               \sqrt{\frac{m_1 m_2 m_3}{m_3-\epsilon}} & 
               m_2-m_1+\epsilon & w(m_2-m_1+\epsilon) \\
               0 & w(m_2-m_1+\epsilon ) &
               m_3-\epsilon  \end{array} \right),
\end{eqnarray}
in which the small parameter $\epsilon$ is related to $w$, i.e.,
$w \simeq \pm \frac{\sqrt{\epsilon m_3}}{m_2}
\left(1+\frac{m_1}{m_2}-\frac{m_2}{2m_3}\right)$,
whose range is determined from the experiments.
Note the sign of $B$ is undetermined from the characteristic
equation but the KM matrix elements are independent of the sign
of $B$.

Then, the diagonalizing matrix $U_L^{l}$ can be written as \cite{others2}
\begin{equation}
U_L^{l} = U_{23}(\theta_{23}) \cdot U_{12}(\theta_{12})
\end{equation}
where
\begin{eqnarray}
U_{12} =\left( \begin{array}{ccc}
              \cos \theta_{12} & \sin \theta_{12} & 0 \\
              -\sin \theta_{12} & \cos \theta_{12} & 0 \\
              0 & 0 & 1 \end{array} \right),
 \qquad
U_{23} =\left( \begin{array}{ccc}
              1 & 0 & 0 \\
              0 & \cos \theta_{23} & \sin \theta_{23}  \\
              0 & -\sin \theta_{23} & \cos \theta_{23} \end{array} \right)
\end{eqnarray}
Since (1,1), (1,3) and (3,1) elements of $M_H$ are zero, 
we may put $U_{13}(\theta_{13})=1$ 
without loss of generality.
The mixing angles $\theta_{12}$ and $\theta_{23}$ can be
written to a very good approximation as
\begin{equation}
\tan \theta_{12}=\sqrt{\frac{m_1}{m_2}}
\end{equation}
and
\begin{equation}
\tan \theta_{23}=\frac{1}{2w}\left[\left(1+\frac{m_1-m_2}{m_3}\right)
-\sqrt{\left(1+\frac{m_1-m_2}{m_3}\right)^2+4w^2\left(\frac{m_1-m_2}
 {m_3}\right)}\right]
\end{equation}

{\it The Case II {\rm with} $K=diag[1, -1, 1]$}:  For a negative $D$, 
the real symmetric  matrix $M_H$ can be diagonalized as 
$U_L^{l}M_H U_L^{l^{\dagger}}
= diag[m_{1}, -m_{2}, m_{3}]$, 
thus reversing the signs of both $m_1$ and $m_{2}$ in Eqs. (2),(5) and (6).
As we noted, a positive $D$ in this case is excluded for the reasons of 
naturalness due to the charged lepton mass hierarchy and
{\it calculability}.

In both cases discussed above,
it turns out that the experimentally allowed range of $w$ in the quark
mass matrices is $0.97 \lesssim  |w| \lesssim 1.87 $ in the leading
approximation \cite{kang2}.  
Thus, we will assign the value of $w$ for the charged lepton sector to
the above range. However, physical observables such as survival and transition 
probabilities for $\nu_{\alpha}$'s are insensitive to the precise value of 
$w$ in the allowed range, as discussed 
in the following.  

\section{Mass matrix for neutrinos}
It is likely that the mass matrix of the charged lepton sector is not 
appropriate for the neutrino sector, since the neutrino oscillation 
experiments do not seem to support such hierarchical pattern for neutrino 
masses as that of quark or charged lepton masses but rather {\it nearly 
degenerate neutrinos} within the three-flavor framework \cite{kkkk,degenerate}.
We will show that such an almost degenerate neutrino scenario
can follow from a neutrino mass matrix, which is clearly different from the 
approach used by others \cite{degenerate}.
We assume that three light neutrinos are Majorana particles.
This can be partly motivated by the fact that there is a dimension-5 operator
which can generate the Majorana masses for SM neutrinos after electroweak
symmetry is spontaneously broken, if one considers the SM as an effective 
field theory of more fundamental theories \cite{eff}.  

In order to construct a neutrino mass matrix so as to be consistent with
the experiments, we consider three observations for neutrinos 
which may be accounted for by assuming massive neutrinos: \\
\begin{itemize}
\item{} solar neutrino data from four different experiments,
the HOMESTAKE \cite{homestake}, GALLEX \cite{gallex}, SAGE \cite{sage},
and the KAMIOKANDE II-III \cite{kamioka}. 
\item{} atmospheric neutrino data measured by four experiments,
the KAMIOKANDE \cite{kamioka2}, Super-Kamiokande \cite{superk},
SOUDAN2 \cite{soudan} and IMB \cite{imb}
\item constraint from the neutrinoless double beta decay experiments 
\cite{doublebeta}
\begin{equation}
\langle m_{\nu_e} \rangle \equiv \left| \sum_{i=1}^3 \eta_i V_{ei}^2 m_i \right|^2 
\le 0.45~{\rm eV}
\end{equation}
where $\eta_i = \pm 1$ depending on the CP property of $\nu_i$.
\item{} the likely need for neutrinos as a candidate of hot dark matter 
\cite{dark}.
\end{itemize}
As is well known, the solar neutrino deficit can be explained through
the MSW mechanism if $\Delta m_{solar}^{2} \approx
6\times 10^{-6}~{\rm eV}^2$ and $\sin^{2} 2\theta_{solar} \approx 7 \times 
10^{-3}$ (small angle case) , or $\Delta m_{solar}^2 \approx 9\times 10^{-6}
~{\rm eV}^2$ and $\sin^{2} 2\theta_{solar} \approx 0.6$ (large angle case)
and through the just-so vacuum oscillations if $\Delta m_{solar}^{2} \approx
10^{-10}~{\rm eV}^2$ and $\sin^{2} 2\theta_{solar} \approx 0.9$ \cite{PDG}.
The atmospheric neutrino problem can be accommodated if $\Delta m_{atmos}^{2} 
\approx 2\times 10^{-3}~{\rm eV}^2$ and $\sin^{2} \theta_{atmos} \approx 1.0$. 
Especially the recent results from CHOOZ and Super-Kamiokande seem to
disfavor  the $\nu_{\mu} \rightarrow 
\nu_{e}$ oscillation as a possible solution to the atmospheric neutrino 
problem \cite{chooz,superk2}. 
So one has to invoke for large mixing between $\nu_{\mu}
\leftrightarrow \nu_{\tau}$.  
If the light neutrinos account for the hot dark matter of the universe, 
one has to require \cite{dark}
\begin{equation}
\sum_{i=1,2,3} | m_{\nu_i} | \lesssim 6~{\rm eV}.
\end{equation}
Thus we see that all three neutrinos may be almost 
degenerate in their masses, with $m_{\nu_i} \lesssim O(1)$~  eV, rather than
$m_{\nu_1} \ll m_{\nu_2} \ll m_{\nu_3}$, as sometimes assumed in the 
three-neutrino mixing scenarios \footnote{The recent LSND data \cite{lsnd1}, 
if confirmed, indicates $\Delta m_{LSND}^{2} \sim 1~{\rm eV}^2$ and $\sin^{2}
\theta_{LSND} \sim 10^{-3}$. Since the conclusions of two different analyses 
\cite{lsnd1,lsnd2} do not agree with each other, we do not consider the 
possibility alluded by the LSND data \cite{lsnd1} in this work.
See, however, Ref.~\cite{kkk} for a discussion when the LSND data is included.
}. 

In this paper, we would like to account for the solar ans atmospheric
neutrino deficits as $\nu_{e} - \nu_{\mu}$ and  $\nu_{\mu} - \nu_{\tau}$
oscillations, respectively.
Once $\nu_{\mu}$ and $\nu_{\tau}$ mixing is taken to be maximal, the
corresponding $2\times2$ mass matrix can be given by
\begin{eqnarray}
\left(\begin{array}{cc}
        A & B \\
        B & A
\end{array} \right).
\end{eqnarray}
After diagonalizing,  we get the eigenvalues $m_{\nu_i} = A\pm B$.
Note that if the parameter $B$ is taken to be small, the atmospheric data of
$\Delta m^2_{atmos}$ can be accommodated.
In addition, $\Delta m^2_{solar}$ can be accommodated by allowing
non-zero (1,1) element of the $3\times3$ mass matrix as follows
\begin{eqnarray}
\left(\begin{array}{ccc}
        C & 0 & 0 \\
        0 & A & B \\
        0 & B & A 
\end{array} \right)
\end{eqnarray}
which can be diagonalized by
\begin{eqnarray}
   U_{\nu} = \left( \begin{array}{ccc}
              1 & 0 & 0 \\
              0 &  \frac{1}{\sqrt{2}} &-\frac{1}{\sqrt{2}}  \\
              0 &   \frac{1}{\sqrt{2}} &\frac{1}{\sqrt{2}} 
              \end{array} \right).
\end{eqnarray}
One can solve for $A, B$ and $C$
by requiring three conditions, $\Delta m_{solar}^2 = 10^{-5}~
{\rm eV}^2$, $\Delta m_{atmos}^2 = 2\times 10^{-3} ~{\rm eV}^2$ 
and Eq.~(7).
Then, the set of parameters $(A, B, C)$ is given by:
\begin{equation}
(A, B, C) = 
(0.33383, 0.001498, 0.3323) ~~(\mbox{eV})
\end{equation}
for which three light neutrinos are almost degenerate with mass around 0.34
eV \footnote{
If $\Delta m^2_{atmos}=10^{-2}~~\mbox{eV}^2$ is used instead, the three
neutrinos are almost degenerate with mass around 1 eV \cite{apctp}.}.
  
\section{Neutrino mixing matrix and predictions}

Combining  the $U^{l}_L$  given by Eq.~(3) with $U_{\nu}$ of Eq.~(11), 
we get the neutrino mixing matrix,
\begin{eqnarray}
V_{\nu} \equiv U_{\nu}^{\dagger} ~U_{L}^{l}
=\frac{1}{\sqrt{2}}\left( \begin{array}{ccc}
               \sqrt{2}c_{12} & \sqrt{2}s_{12} & 0 \\
                s_{12}(-c_{23}+s_{23}) &
                c_{12}(c_{23}-s_{23}) &
               s_{23}+c_{23} \\
                s_{12}(c_{23}+s_{23}) &
                -c_{12}(c_{23}+s_{23}) &
               -s_{23}+c_{23} 
                \end{array} \right),
\end{eqnarray}
where we have abbreviated $\cos\theta_{ij}$ and $\sin\theta_{ij}$ as
$c_{ij}$ and $s_{ij}$ respectively.
We note that the mixing matrix is independent of neutrino masses, 
although it depends on the charged lepton masses.
For the whole range of $0.97 \lesssim |w| \lesssim 1.87 $, 
the neutrino mixing matrix is given by
\begin{eqnarray}
    |V_{\nu}| = \left( \begin{array}{ccc}
               0.9952 & 0.0692 & 0.0  \\
               0.0453 & 0.6520 & 0.7531 \\
               0.052 & 0.7947 & 0.6551
          \end{array} \right) \sim
     \left( \begin{array}{ccc}
               0.9952 & 0.0692 & 0.0  \\
               0.0440 & 0.6326 & 0.7307 \\
               0.0506 & 0.7271 & 0.6356
          \end{array} \right).
\end{eqnarray}
Note that our lepton mixing matrix {\it predicts} zero for $(V_{\nu})_{13}$ element,
i.e., the $\nu_e$-tau coupling is forbidden,
which makes $\nu_{e} \leftrightarrow \nu_{\mu}$ and 
$\nu_{\mu} \leftrightarrow \nu_{\tau}$ oscillations to be effectively
a two-channel problem.

Now, we check if the solutions of three neutrino mass eigenvalues satisfy 
the constraint coming from the neutrinoless double $\beta-$decay,
as well as other data from neutrino oscillation experiments.   
The neutrino mixing matrix Eq.~(13) and neutrino mass eigenvalues 
lead to
\begin{equation}
\langle m_{\nu_e} \rangle \simeq 0.33 ~{\rm eV}
\end{equation}
for $w=0.97-1.87$. All of these solutions  are well below the current upper 
limit given in Eq.~(7). If we begin to increase the neutrino masses in order 
to make it dominant hot dark matter candidates, we cease to satisfy the 
$(\beta\beta)_{\nu 0}$ constraint, Eq.~(7).

Next, we study the transition and survival probabilities of the neutrinos.
In order to calculate the transition probabilities, the mass differences
$\Delta m^2_{ij}=m^2_{\nu_i}-m^2_{\nu_j}$ should be identified 
with $\Delta m^2_{solar}$ or $\Delta m^2_{atmos}$.
Among the possibilities, it turns out that only the case for
$\Delta m^2_{solar}=\Delta m^2_{12}$ and $\Delta m^2_{atmos}=\Delta m^2_{23}$
can fit the available data quite well, and thus we will consider 
henceforth only this  case.
%
In particular, we find that the probability
$P(\nu_{e}\rightarrow \nu_{\mu})$ and $P(\nu_{\mu}\rightarrow \nu_{\tau})$
is  changed up to about $10\%$
with the value of $w$.

Further test of our ansatz is provided with the long baseline experiments
searching for $\nu_{\mu} \rightarrow \nu_{\tau}$ oscillation in the range 
of  $\Delta m_{\mu\tau}^2 \simeq 10^{-3}~{\rm eV}^2$. 
The MINOS \cite{minos} and K2K \cite{k2k} sensitivities to $\Delta m^2$
at $90\%$ CL can go down to $\Delta m^2 = 1.2 \times 
10^{-3}~\mbox{eV}^2$ and $2.0\times 10^{-3}~\mbox{eV}^2$, respectively,
while the ICARUS \cite{ica} sensitivity is achieved at $\Delta m^2=3.0\times 10^{-3}~
\mbox{eV}^2$.
 Our prediction is that 
\begin{equation}
\sin^2 {2\theta_{\mu \tau}} \simeq 0.86 - 0.97
\end{equation}
with $\Delta m_{\mu\tau}^2 = 2\times 10^{-3}~~{\rm eV}^2$ 
for the allowed range of $w$. 
This can be tested at the MINOS and K2K experiments searching for the
$\nu_{\mu} \rightarrow \nu_{\tau}$ oscillations in the foreseeable future,
but is beyond the sensitivity to $\Delta m^2$ at $90\%$ CL being achieved at ICARUS.
Future experiment on the $\nu_{\mu} \leftrightarrow \nu_{\tau}$ oscillation
from the MINOS and K2K will exclude our model for charged lepton and neutrino
mass matrices.

\vspace{.3in}

\section{Conclusion}

In conclusion, we investigated in this paper phenomenological consequences of 
the lepton mass matrix ansatzs with the minimal number of parameters, three 
each in the charged lepton  
and Majorana neutrino mass matrices 
We find the 
ansatze Eqs.~(1) and (10) lead to a lepton mixing matrix which is
consistent with the current data on various types of neutrino oscillation 
experiments except the controversial LSND data.
Three light Majorana neutrinos can constitute a part of hot dark matter, 
with $\Sigma | m_{\nu_i} | \sim 1~$ eV without contradicting the 
constraint from  neutrinoless beta decay experiments.  The resulting
amplitude of
$\nu_{\mu} \leftrightarrow \nu_{\tau}$ oscillation $\sin^2{2\theta_{\mu \tau}}$
is $0.86 - 0.97$
with $\Delta m_{\mu\tau}^2 = 2\times 10^{-3}~{\rm eV}^2$ for the range of
$w$ under consideration,  which is
consistent with the atmospheric neutrino oscillation and will be  
tested at the MINOS and K2K experiments. Finally, three neutrinos 
being  almost degenerate, we expect that the lepton family number
breaking effects in  $\mu \rightarrow e \gamma$ and $\mu
\rightarrow 3 e$ and analogous tau decays will be very small.

\acknowledgements   
We would like to thank Soo Bong Kim for valuable informations on the long
base line experiments.
Two of us (SKK,JEK) would like to thank the members of the High Energy Theory 
Group for the warm hospitality extended to them at Brown University.
This work is supported in part by the 
KOSEF Postdoctoral fellowship(SKK), the KOSEF through  Center for Theoretical 
Physics at Seoul National University and by KOSEF Contract No. 971-0201-002-2
(PK), SNU-Brown Exchange Program (KK, JEK, PK), the Ministry of Education 
through the Basic Science Research Institute, Contract No. BSRI-97-2418 (JEK,
PK), Hoam Foundation (JEK), Distinguished Scholar Exchange Program of Korea 
Research Foundation(JEK,PK), and also 
the US DOE Contract DE-FG-02-91ER40688 - Task A (KK).


\begin{references}
\bibitem{kkkk} K. Kang, S. K. Kang, J. E. Kim and P. Ko, Mod. Phys. Lett. A
               {\bf 12}, 553 (1997), Report No. hep-ph/9611369.
\bibitem{degenerate} D. Caldwell and R. N. Mohapatra, Phys. Rev. D{\bf  48},
      3259 (1993); {\it ibid.}{\bf 50}, 3477 (1994); 
S. T. Petcov and A. Y. Smirnov,
      Phys. Lett. B{\bf 322}, 109 (1994); A. Joshipura, Z. Phys. C{\bf 64},
31 (1994); Phys. Rev. D{\bf 51}, 1321 (1995); 
A. Ioannissyan and J. W. F. Valle,
 Phys. Lett. B{\bf 332}, 93 (1994); D. G. Lee and R. N. Mohapatra, {\it ibid.}
 {\bf 329}, 463 (1994); C. Y. Cardall and G. M. Fuller, Nucl. Phys. Proc. Supp.
{\bf 51B}, 259 (1996);
 A. Acker and S. Pakvasa, Phys. Lett. B{\bf 397}, 209 (1997).
\bibitem{superk2} Y. Fukuda et al., the Super-Kamiokande Collaboration,
reported the two-flavor $\nu_{\mu}\leftrightarrow \nu_{\tau}$ oscillations
with $\sin^2{2\theta} > 0.82$ and $5\times 10^{-4} < \Delta m^2 < 6 \times
10^{-3}~~ \mbox{eV}^2$ at $90\%$ confidence level.
\bibitem{wein} S. Weinberg, ``The Problem of Mass", in a Festschrift for
I.I.Rabi, L. Motz, ed., (N.Y.Academy of Science, N.Y.1977).
See also F. Wilczek and A. Zee, Phys. Lett. B{\bf 70}, 418 (1977); {\it ibid.}
{\bf 72}, 504 (1978).
\bibitem{fritz}H. Fritzsch, Phys. Lett. B{\bf 73}, 317 (1978);
Nucl. Phys. B{\bf 155},189 (1979).
\bibitem{kang1} K. Kang and A. C. Rothman 
             , Phys. Rev. Lett. {\bf 43}, 1548 (1979); Phys. Rev. D
             {\bf 24}, 167 (1981).
\bibitem{kangt}K. Kang and S. Hadjitheodoridis, Phys. Lett. B{\bf 193},
              504 (1987); K. Kang, J. Flanz, and E. Paschos, Z. Phys.
              C {\bf 55}, 75 (1992). See also H. Harari and and Y. Nir,
              Phys. Lett. B{\bf 195}, 586 (1987).
\bibitem{kang2} K. Kang and S. K. Kang, Phys. Rev. D{\bf 56}, 1511 (1997),
    Report No. hep-ph/9704253.
\bibitem{others1} P. Kaus and S. Meshkov, Phys. Rev. D{\bf 42}, 1863 (1990);
        H. Fritzsch and D. Holtmannsp\"{o}tter, Phys. Lett. B{\bf 338},
              290 (1994); H. Fritzsch and Z. Xing, Phys. Lett. B
              {\bf 353}, 114 (1995); H. Fritzsch, 
              in {\it Rencontres de Moriond}, Les Arcs, France
              (Electroweak Interactions), Mar. 1996, Report No. hep-ph/9605388;
              P.S.Gill and M. Gupta, J. Phys.G{\bf 21}, 1 (1995); Int. J.
              Mod. Phys. A{\bf 11}, 4805 (1996);
              H. Lehmann, C. Newton, and
              T. T. Wu, Phys. Lett. B{\bf 384}, 249 (1996);
              Z. Xing, Report No. DPNU-96-39, hep-ph/9609204;
              P.S.Gill and M. Gupta, Phys. Rev. D{\bf 56}, 3143 (1997) and
              references therein.
\bibitem{ramond}P. Ramond, R. G. Roberts and G. G. Ross, Nucl. Phys. B
              {\bf 406}, 19 (1993).
\bibitem{homestake}R. Davis et al., Proc. 21st Int. Cosmic Ray Conf. Univ.
   of Adelaide, ed. R. J. Protheroe, Vol. 12, 143 (1990).
\bibitem{gallex} P. Anselmann {\it et al.}, Phys. Lett. B{\bf 285}, 376 (1992)
{\it ibid.} {\bf 285}, 390 (1992);
{\it ibid.} {\bf 314}, 445 (1993); {\it ibid.} {\bf 327}, 377 (1994).
\bibitem{sage} A.I. Abazov {\it et al.},  Phys. Rev. Lett. {\bf 67}, 3332 
(1991).
\bibitem{kamioka} K.S. Hirata {\it et al.},  Phys. Rev. Lett. {\bf 65}, 1297
 (1990) ;  {\it ibid.} {\bf 66}, 9 (1991) ;  Phys. Rev. D{\bf 44}, 2241 (1991).
\bibitem{kamioka2} K. S. Hirata {\it et al.}, Phys. Lett. B{\bf 205}, 416 (1988) ; {\it ibid.} {\bf 280}, 146 (1988); Y. Fukuda {\it et al.}, {\it ibid.} 
{\bf 335}, 237 (1994).
\bibitem{superk} E. Kearns, Talk at {\it News about SNUS}, ITP Workshop,
 Santa Barbara, Dec. 1997.
\bibitem{soudan} P.J. Litchfield, Proc. Int. Europhys. Conference on High 
Energy Physics, Marseille, 557 (1993). (Editions Frontieres, ed., J. Carr 
and M. Perrottet.)
\bibitem{imb} R. Becker-Szendy {\it et al.}, Phys. Rev. D{\bf 46}, 3720 (1992);
D. Casper  {\it et al.}, Phys. Rev. Lett. {\bf 66}, 2561 (1992).
\bibitem{msw} L. Wolfenstein, Phys. Rev. D{\bf 17}, 2369 (1978);
S. P. Mikheyev and A. Yu. Smirnov, Yad. Fiz. {\bf 42}, 1441 (1985);
Sov. J. Nucl. Phys. {\bf 42}, 913 (1986).
\bibitem{chooz} Apollonio {\it et al.}, CHOOZ Collaboration, hep-ex/9711002.
\bibitem{kang3}A. C. Rothman and K. Kang (Ref. 2); K. Kang in {\it
             Particles and Fields 2}, ed. A. Z. Capri and A. N. Kamal
             (Plenum Pub. Co., 1983); K. Kang and S. Hadjitheodoridis,
              in Proc. of the 4th Symp. Theor. Phys., ed. H. -S. Song
             (Mt. Sorak, Korea, 1985).
\bibitem{others2}H. Fritzsch and Z. Xing, Phys. Lett. {\bf B 413}, 396 (1997);
               H. Lehmann, C. Newton, and T. T. Wu, Ref. \cite{others1}.
\bibitem{eff} R. Barbieri, J. Ellis and M.K. Gaillard, Phys. Lett. 
              {\bf B 90}, 249 (1980).
\bibitem{doublebeta} Heidelberg-Moscow Collab., Phys. Lett. \textbf{B 407} 
219 (1997) ; H.V. Klapdor-Kleingrothaus {\it et al.}, hep-ph/9712381.
\bibitem{dark}G. Gelmini and E. Roulet, Rep. Prog. Phys. {\bf 58}, 1207 (1995).
\bibitem{PDG} Particle data Group, R.M.Barnett {\it et al.}, Phys Rev. D{\bf 54}, 1 (1996).
\bibitem{lsnd1} C. Athanassopoulos {\it et al.}, Phys. Rev. Lett. {\bf 75}, 
2650 (1995). 
\bibitem{lsnd2} J.E.  Hill, Phys. Rev. Lett. {\bf 75},  2654 (1995).
\bibitem{kkk} K. Kang, J. E. Kim and P. Ko, Z. Phys. C{\bf 72}, 671 (1996).
\bibitem{apctp} K. Kang and S. K. Kang, hep-ph/9802328,
               talk given at the 
               workshop Pacific Particle Physics Phenomenology, 
             Seoul National University,
      Seoul, Korea, 31 October-2 November, 1997.
\bibitem{minos} Minos collaboration, {\it P-875:  A long baseline neutrino 
               oscillation experiment at Fermilab}, NuMI-L-63, Feb. 1995.
\bibitem{k2k} M. Sakuda, K2K collaboration, {\it The KEK-PS Long Baseline 
              Neutrino Oscillation Experiment(E362)},  talk given at the 
               workshop Pacific Particle Physics Phenomenology, 
             Seoul National University,
      Seoul, Korea, 31 October-2 November, 1997.
\bibitem{ica} A. Rubbia, Nucl. Phys. B (Proc. Suppl.) {\bf 66}, 436 (1998).
\end{references}
\end{document}